\documentclass{article}

\usepackage{graphicx}
\graphicspath{{./}}
\usepackage{dcolumn}
\usepackage{bm}
\usepackage[mathlines]{lineno}
\usepackage{float}
\usepackage{amsmath}
\usepackage{braket}
\renewcommand{\Im}{\mathrm{Im}}
\newcommand{\Tr}{\mathrm{Tr}}
\newcommand{\ii}{\mathrm{i}}
\newcommand{\dd}{\mathrm{d}}
\newcommand{\ave}[1]{\braket{#1}_{\mathrm{eq}}}
\newcommand{\EF}{E_{\mathrm{F}}}

\usepackage{listings} 
\usepackage[backend=biber, style=numeric]{biblatex}
\title{High-performance Computation of Kubo Formula with Vectorization of Batched Linear Algebra Operation}
\author{
  Yuta Yahagi\thanks{Secure System Platform Laboratory, NEC Corporation},\and
  Toshihiro Kato\thanks{HPC Department, NEC Corporation}
  }
\date{July 27,2023}

\begin{document}
\maketitle

\begin{abstract}
  We have proposed a method to accelerate the computation of Kubo formula optimized to vector processors. 
  The key concept is parallel evaluation of multiple integration points, enabled by batched linear algebra operations.
  Through benchmark comparisons between the vector-based NEC SX-Aurora TSUBASA and the scalar-based Xeon machines in node performance, 
  we verified that the vectorized implementation was speeded up to approximately 2.2 times faster than the baseline. 
  We have also shown that the performance improvement due to padding, indicating that avoiding the memory-bank conflict is critically important in this type of task.
\end{abstract}

\section{Introduction}

One of the main challenges in data-driven material development is the small data problem. 
It is not uncommon for data scientific methodologies to fail to work due to a lack of sufficient data in a target domain.
For overcoming this problem, computational approaches based on high-throughput first-principles calculations have attracted much attention and are actively used in data generation \cite{Jain2011, Curtarolo2012, Curtarolo2013, Yang2018}. 
Compared with experimental approaches, the computational approach has advantages in terms of data production, allowing us to construct a large-scale dataset or to run an exhaustive virtual screening.

Currently, almost all high-throughput calculation projects stand on the Kohn--Sham density functional theory (KS-DFT), which is an established and generic method for predicting the electronic structure of materials without assuming empirical parameters. 
The DFT fundamentally characterizes the ground state of electrons.
It enables the computation of static quantities such as electron density, total energy, force fields, and magnetic moments, thus providing useful insights into structural stability, chemical reactions, and various practical applications\cite{Martin2004}.

External-field responses such as transport properties and optical properties not only give helpful knowledge of materials under a driving forces but also allow direct comparison between calculations and measurements.
However, unlike the static quantities, these properties are not included in the standard DFT procedure because they must gather electron excitation, which is out of the range of DFT.
Computing response functions in the first-principles level, we have to consider some theoretical extensions of DFT such as time-dependent DFT or additional post-process calculation depending on specific tasks.

Recently, a post-process scheme based on the Kubo formula with an effective tight-binding model has been widely used in the context of high-throughput calculations. 
The Kubo formula is a general formula for linear response phenomena, which can be evaluated in a single framework regardless of the system and target \cite{Kubo1957}. 
Usually, the electronic structure obtained by DFT is often downfolded to a tight-binding (TB) model such as Wannier TB to make its evaluation easy \cite{Marzari2012}. 
On the basis of such a scheme, several studies have successfully produced datasets of transport properties, for example, the charge conductivity, the thermoelectric conductivity, and the spin conductivity\cite{Sakai2020,Zhang2021a,Samathrakis2022,Zelezny2023}.

Although the Kubo formula is an attractive method for high-throughput calculations due to its versatility and stability, it is computationally expensive, limiting the yield of workflow even if combined with TB models. Roughly estimated, the evaluation of Kubo formula typically costs 10 times as much as that of DFT calculation.
Moreover, in contrast to the DFT core program, for which implementations utilizing a general-purpose graphics processing unit (GPGPU) exist, speeding up of Kubo formula remains a major challenge because it involves overcoming multiple computational bottlenecks.
Modern high-performance computing (HPC) implementations of the Kubo formula can play a vital role in increasing the yield of data generation.

In this study, we develop an efficient method of evaluating the Kubo formula by optimizing for vector processors. As a practical example, we perform benchmark tests in computation of the anomalous Hall conductivity (AHC) \cite{Nagaosa2010} and show the results of performance comparisons between the vectorized method and an existing method of CPU implementation.

\section{Numerical method}

\subsection{Kubo formula}

Within the framework of the Kubo formula, a linear response of an observable $\hat{B}$ can be generally expressed as
\begin{align}
  \label{eq:KuboFormula}
  \chi_{BA}(\omega)=\frac{\ii}{\hbar V}\int_{0}^{\infty} \dd t e^{\ii \omega t}
  \ave{[\hat{B}(t),\hat{A}(0)]},
\end{align}
where $V$ is the volume, the hatted symbols are operators, and $\chi_{BA}(\omega)$ is the response function of an observable $\hat{B}$ under an external field $F(t)=Fe^{\ii \omega t}$ giving a perturbation $-\hat{A}F(t)$. 
$\ave{\cdots}$ is an equilibrium average with respect to the non perturbative Hamiltonian.
Considering a single-electron system, this average can be reduced to a Brillouin-zone integral $\ave{\hat{X}} \to (2\pi)^{-3}V \int_{\mathrm{B.Z.}}\dd \bm{k} \Tr \hat{X}(\bm{k})$ for a periodic system, otherwise a real-space integral $\ave{\hat{X}} \to \int \dd \bm{r} \Tr \hat{X}(\bm{r})$.

From the perspective of numerical calculation, Eq.(\ref{eq:KuboFormula}) should consist of a three-dimensional integration whose integrand is the trace of a matrix resulting from multiple linear algebra operations.
The integrand, depending on a specification, mostly conducts matrix products with either a full-diagonalization or a matrix inverse.
Eventually the computational task involves a nested loop for the numerical integration and the matrix oprations, hence the time complexity is approximately $O(N_k*N_d^3)$ where $N_k$ is the number of integration points and $N_d$ is the dimension of the matrix corresponding to the number of single-electron bands. 

Here, let us focus on the intrinsic AHC, $\sigma_{yx}$, in a periodic, single-electron system at zero temperature for demonstration.
Eq.(\ref{eq:KuboFormula}) can be rewritten as
\begin{align}
  \label{eq:AHC_sum}
  \sigma_{yx}&=\sum_{\bm{k}}^{N_k}f(\bm{k}), \\
  \label{eq:AHC_integrand}
  f(\bm{k})&=\frac{\hbar}{\pi}\sum_{n}^{N_d}\sum_{m\neq n}^{N_d}
  \Im \frac{[\hat{J}_y(\bm{k})]_{nm} [\hat{J}_x]_{mn}(\bm{k})}
    {(E_{m}(\bm{k}) -E_{n}(\bm{k}))^2}\theta(\EF - E_{n}(\bm{k})),
\end{align}
where $\hat{J}_{\alpha}(\bm{k})$ is a charge current operator, 
$E_{n}(\bm{k})$ is an eigenvalue, 
$\EF$ is the Fermi energy, 
and $\theta(x)$ is a unit step function.
In this formula, all operators are represented in $N_d$-dimensional matrices.
Note that Eq.(\ref{eq:AHC_integrand}) is valid for a periodic system, whereas for a system with impurities, we have to compute Green's functions, i.e., matrix inverse, instead of eigenvalues.

\subsection{Baseline implementation}

A minimal code evaluating Eqs.(\ref{eq:AHC_sum},\ref{eq:AHC_integrand}) is summarized as the pseudo code in List\ref{code:Baseline}, which consists of a $\bm{k}$ integral of (1) construction of matrices $\hat{H}(\bm{k})$ and $\hat{J}_{\alpha}(\bm{k})$, (2) full-diagonalization of $\hat{H}(\bm{k})$, (3) conversion of $\hat{J}_{\alpha}(\bm{k})$ into the eigenbasis, and (4) evaluation of Eq.(\ref{eq:AHC_integrand}). The integral requires $N_k$ times of the evaluations of (1)-(4), which takes $O(N_d^3)$, resulting in the total time complexity $O(N_kN_d^3)$. In particular, the most time consuming part is the full-diagonalization (2). 

To accelerate the calculation, a parallel programming technique (for instance, MPI parallelization) can be applied to either the $\bm{k}$ integral or the matrix calculations.
In high-throughput computation, the former is more effective than the latter.
This is because the parallelization of matrix calculations is not effective in many cases where the material is composed of several chemical species per unit cell, in which case the dimension of the matrix $N_d$ is also smaller. 
Moreover, the integrand is spiky in the momentum space and requires a very fine computational mesh of approximately $10^6-10^7$ points, incurring a critical amount of overhead when MPI is used inside of the integral. 

\subsection{Vectorized implementation}
\label{sec:Vectorized}

\begin{figure}[H]
  \centering
  \includegraphics[width=0.9\linewidth]{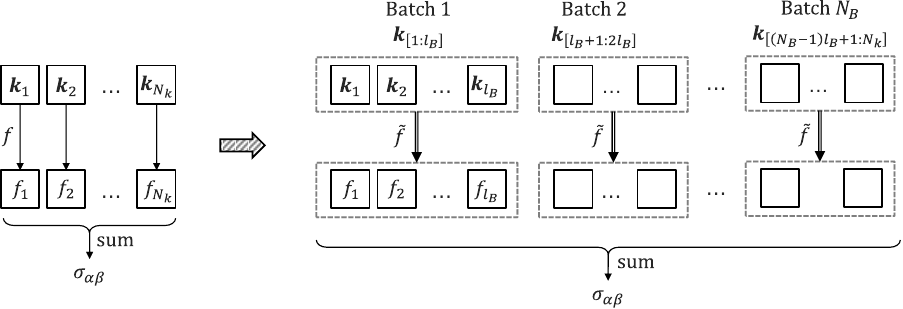}
  \caption{Schematics of (left) the baseline implementation and (right) the vectorized implementation of the Kubo formula. Here, $f$ is a function of $\bm{k}$ and $\tilde{f}$ is a function of a list of $\bm{k}$, corresponding to those in Eq.(\ref{eq:batched_sum}), respectively.}
  \label{fig:Batched_Kubo_formula}
\end{figure}

In this section, we propose an alternative, namely vectorized implementation of the Kubo formula, which is optimized for vector processors.
The key idea is batch evaluation of multiple integration points, as depicted in Fig.\ref{fig:Batched_Kubo_formula}.
Let $N_B$ be the number of batches and $l_B$ be the batch size taken to be the same as the vector length. Here, $N_k=N_B*l_B$ holds.
We first split the $\bm{k}$-integral loop of the baseline code into two loops: an inner one for the batches of length $l_B$ and an outer one for the sum of them, that is,
\begin{align}
  \label{eq:batched_sum}
  \sigma_{yx}=\sum_{\bm{k}}^{N_k}f(\bm{k}) 
  \xrightarrow{\mathrm{Batched}}
  \sum_{b=1}^{N_B}\sum_{l=1}^{l_B}f(\bm{k}_{bl_B+l})
  \equiv \sum_{b=1}^{N_B}\tilde{f}(\bm{k}_{[(b-1) l_B : b l_B]}),
\end{align}
where $\bm{k}_{[n:m]} \equiv [\bm{k}_n,\bm{k}_{n+1}\dots,\bm{k}_m]$.
In other words, a part of original loop is pushed into the vectorized integrand $\tilde{f}$. 
The performance of vectorization becomes apparent when the loop is sufficiently longer than $l_B$. In our case where $l_B \ll N_k$ is well satisfied, we can expect a sufficient benefit of vectorization.

List\ref{code:Vectorized} represents a pseudo code of this implementation.
Compared with List\ref{code:Baseline}, the procedure is exactly equivalent, but the loop cycles are executed per batch, and each matrix operation is performed in a batch. 
Here, we prepare a batched version of the matrix product and the diagonalization; the latter is obtained by modifying the EISPACK subroutine \cite{Smith1976}.
Such a batched linear algebra operation scheme is known as the Batched-BLAS in recent literature \cite{Dongarra2017a, Dongarra2017b, Abdelfattah2021}.

In addition, to avoid degradation due to memory-bank conflict, the matrix size is adjusted by padding.
The benefits of the padding will be demonstrated in the next section.

To end this section, we comment on the case of a large-scale system involving hundreds of atoms.
In such a case, the dimension of the matrix $N_d$ is sufficiently large, and parallelization of the matrix operations becomes effective.
Because the required $N_k$ is in a tradeoff relationship between the unit cell size, we can choose the best strategy of parallelization depending on the problem.
As large-scale systems are beyond the scope of this paper, we will not discuss them further.

\section{Benchmark results}

To examine the performance of the vector implementation, we conduct a demonstration on a vector machine NEC SX-Aurora TSUBASA where the vector length is 256 \cite{Aurora}.
For comparison, the baseline implementation is also tested on a CPU machine, Xeon Gold, with 32-process parallelization by flat MPI.
We are aware that comparison between different architectures is not straightforward, but here we compare single node performances as a compromise measure.
The computing and compiling environments are listed in Table \ref{table:Environment}.

In this demonstration, we measure the computation times to evaluate AHC of the bcc-Fe and its supercell structures. 
Wannier TB models are constructed from the DFT calculation using the \textrm{Quantum Espresso} code \cite{Giannozzi2009} and the \textrm{Wannier90} code \cite{Pizzi2020}.
We refer to the \textrm{WANNIER-LINEAR-RESPONSE} code \cite{JZLinres} as the baseline implementation of Kubo formula and vectorize it by the method introduced in Sec.\ref{sec:Vectorized}.

Let us begin with the vector processing rate, a performance indicator defined as the proportion of vectorized processing in total processing.
According to the profiler, our vectorized code achieves sufficiently high scores reaching 98-99 \% in all demonstrations.

\begin{figure}[H]
  \centering
  \includegraphics[width=0.8\linewidth]{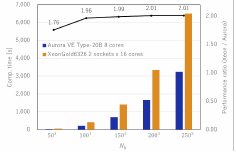}
  \caption{
    (Color online) Computation times on the Xeon and the SX-Aurora versus the number of $\bm{k}$-points, $N_k$. Their performance ratio is defined as (Xeon computing time)/(SX-Aurora computing time). Here, the [1,1,2]-supercell of bcc-Fe is used.
  }
  \label{fig:Nk_dependence}
\end{figure}

Fig.\ref{fig:Nk_dependence} shows the computation times as a function of $N_k$. Here, we use the [1,1,2]-supercell where $N_d=48$ as it has 4 atoms with 2 spins and 6 Wannier projections per atom. From this data, we can see that the results on the SX-Aurora are about twice as fast as those of 32-process parallel execution on Xeon Gold, indicating a significant acceleration due to the vectorized implementation. 
This trend becomes more pronounced as $N_k$ increases.
We again emphasize that this comparison is only a single aspect and does not lead general superiority of the vector processor over the CPU.
Rather, this is clear evidence of the usefulness of the vectorization technique on this task.

\begin{figure}[H]
  \centering
  \includegraphics[width=0.8\linewidth]{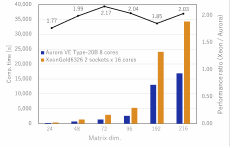}
  \caption{
    (Color online) Computation times on the Xeon and the SX-Aurora versus the dimension of matrix, $N_d$. Their performance ratio is defined the same as in Fig.\ref{fig:Nk_dependence}. Here, $N_k=150^3$ is used.
  }
  \label{fig:Nd_dependence}
\end{figure}

Next, Fig.\ref{fig:Nd_dependence} shows the computation times on the $N_k=150^3$ case as a function of $N_d$.
Although the performance ratios varied by matrix size, in all cases, the SX-Aurora is more than 1.7 times faster than the Xeon Gold and 2.17 times at maximum.
Note that the performance ratio on $N_d=192$ is lower because it is a special condition for better performance of the Xeon Gold rather than defects of the SX-Aurora. 
Together with the results in Fig.\ref{fig:Nk_dependence}, we claim that vectorized implementations have shown significant speedups in a wide range of tasks on small-scale systems, which is a desirable feature in high-throughput computations. 

\begin{figure}[H]
  \centering
  \includegraphics[width=0.8\linewidth]{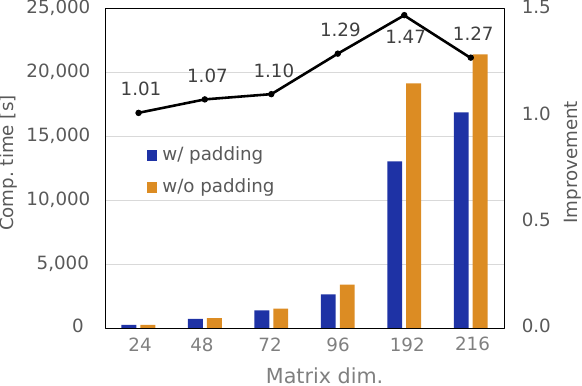}
  \caption{
    (Color online) Computation times on the SX-Aurora with and without the padding versus the dimension of matrix, $N_d$. 
    Improvement ratio is defined as (computing time without padding)/(computing time with padding).
    Here, $N_k=150^3$ is used.
  }
  \label{fig:Padding_dependence}
\end{figure}

Finally, we examine the effect of the padding on the performance. 
Fig.\ref{fig:Padding_dependence} compares the vectorized implementation with and without the padding, showing that the padding improves the performance by 1.5 times at maximum.
The improvement becomes notable where $N_d \ge 96$ (8 atoms), that is, the memory-bank conflict causes serious performance degradation except in very small systems. 
Hence, padding is critically important in practical use.

\section{Conclusion}

In this study, we have proposed a method to accelerate the computation of Kubo formula by using batched matrix operations optimized to vector processors. Our findings have led to the following significant insights:

First, the introduction of vectorization was demonstrated to accelerate the computation of Kubo formula. 
Through benchmark comparisons between the vector-based NEC SX-Aurora TSUBASA and the scalar-based Xeon machines, we verified that the SX-Aurora has up to approximately 2.2 times faster node performance. 
This result indicates that our proposed method is effective and leads to increased data production in the context of high-throughput calculations.

The computational performance improvement due to padding was observed. This result underscores the importance of avoiding the memory-bank conflict to improve computational efficiency.

Furthermore, our approach is not limited to Kubo formula; it can be extended to various similar tasks relying on batched linear algebra operations.
The vectorization approach could become a powerful alternative to the MPI in such cases.
In addition, the same strategy can be applied to GPGPU as long as memory consumption is acceptable. 
The GPGPU implementation is an interesting future work.

\section*{Aknowledgement}
Y.Y and T.K. are grateful to Dr. Taizo Shibuya at NEC Corporation for his valuable suggestions.

\section*{Contents}

\lstset{boxpos=H}
\begin{lstlisting}[caption={Baseline implementation}, label={code:Baseline},escapechar=@]
! @$\bm{k}$@-loop (MPI parallel may be applied here)
for @$\bm{k}$@ in {@$\bm{k}_1, \bm{k}_2,\dots, \bm{k}_{N_k}$@}
  ! Define Hamiltonian and current operators
  @$\hat{H}, \hat{J}_{\alpha}, \hat{J}_{\beta}$@ = define_matrices(@$\bm{k}$@)
  ! Diagonalization of Hamiltonian 
  ! (@$\bm{E}$@ : Eigenvalues, @$\hat{U}$@ : Eigenvectors)
  @$\bm{E}, \hat{U}$@ = diagonalize(@$\hat{H}$@)
  ! Convert @$\hat{J}_{\alpha}$@ into the eigenbasis as @$\hat{J}'=\hat{U}\hat{J}\hat{U}^{\dagger}$@
  @$\hat{J}'_{\alpha}$@ = convert(@$\hat{J}_{\alpha}, \hat{U}$@); @$\hat{J}'_{\beta}$@ = convert(@$\hat{J}_{\beta}, \hat{U}$@)
  ! Evaluate AHC by Eq.(@\ref{eq:AHC_integrand}@)
  @$\sigma_{\alpha\beta}$@ += evaluate(@$\bm{E}, \hat{J}'_{\alpha}, \hat{J}'_{\beta}$@)
end for
\end{lstlisting}

\begin{lstlisting}[caption={Vectorized implementation}, label={code:Vectorized},escapechar=@]
  ! Separate @$\bm{k}$@-loop by batches of length @$l_B$@
  for @$\bm{k}[:]$@ in {@$\bm{k}_{[1:l_B]}, \bm{k}_{[l_B+1:2l_B]},\dots, \bm{k}_{[(N_B-1)l_B+1:N_k]}$@}
    ! Define lists of Hamiltonian and current operators
    @$\hat{H}[:], \hat{J}_{\alpha}[:], \hat{J}_{\beta}[:]$@ = define_matrices(@$\bm{k}[:]$@)
    ! Batched diagonalization of Hamiltonians
    @$\bm{E}[:], \hat{U}[:]$@ = diagonalize(@$\hat{H}[:]$@)
    ! Convert @$\hat{J}_{\alpha}[:]$@ into the eigenbasis by batched matrix product
    @$\hat{J}'_{\alpha}[:]$@ = convert(@$\hat{J}_{\alpha}[:], \hat{U}[:]$@); @$\hat{J}'_{\beta}[:]$@ = convert(@$\hat{J}_{\beta}[:], \hat{U}[:]$@)
    ! Evaluate AHC by Eq.(@\ref{eq:AHC_integrand}@) and aggregate the batch
    @$\sigma_{\alpha\beta}$@ += evaluate(@$\bm{E}[:], \hat{J}'_{\alpha}[:], \hat{J}'_{\beta}[:]$@)
  end for
\end{lstlisting}

\begin{table}[H]
  \caption{Computing and compiling environments}
  \label{table:Environment}
  \centering
  \begin{tabular}{lcc}
    \hline
     & Vector machine & Scalar machine \\
    \hline \hline
    Hardware (1 node) & VE Type-20B 8 cores \cite{VEType20} & Xeon Gold 6362 2 sockets * 16 cores \\
    Compiler & NEC SDK 4.0.0 \cite{NECSDK} & Intel ifort 19.1.2.254 \\
    MPI & NEC MPI 3.0.0 & Intel MPI ver. 2019 \\
    Numerical library & NLC 2.3 & Intel MKL \\
    \hline
  \end{tabular}
\end{table}

\newrefcontext[sorting=count]
\printbibliography[title=References]

\end{document}